\documentclass[times,twocolumn,final]{elsarticle}

\usepackage{medima}
\usepackage{framed,multirow}

\usepackage{amssymb}
\usepackage{latexsym}

\usepackage{url}
\usepackage{xcolor}

\usepackage{hyperref}

\usepackage{enumitem}
\usepackage{amsmath}
\usepackage{float}
\usepackage{caption}
\usepackage{subcaption}
\usepackage{array}
\newcolumntype{P}[1]{>{\centering\arraybackslash}p{#1}}
\newcolumntype{M}[1]{>{\centering\arraybackslash}m{#1}}
\usepackage{makecell}
\usepackage{dblfloatfix}

\definecolor{newcolor}{rgb}{.8,.349,.1}

\journal{XXX}

\begin{document}

\verso{Wei Dai \textit{et~al.}}

\begin{frontmatter}

\title{Deeply Supervised Skin Lesions Diagnosis with Stage and Branch Attention}

\author[1]{Wei Dai}
\author[1]{Rui Liu}
\author[1]{Tianyi Wu}
\author[1]{Min Wang}
\author[2]{Jianqin Yin}
\author[1,3]{Jun Liu\corref{cor1}}
\cortext[cor1]{Corresponding author.}
\ead{jun.liu@cityu.edu.hk}

\address[1]{ City University of Hong Kong, Hong Kong, China}
\address[2]{Beijing University of Posts and Telecommunications, Beijing, China}
\address[3]{Shenzhen Research Institute, City University of Hong Kong, Shenzhen, China}

\received{xx June 2022}
\finalform{xx xx 2022}
\accepted{xx xx 2022}
\availableonline{xx xx 2022}
\communicated{xxx xxx}

\begin{abstract}
Accurate and unbiased examinations of skin lesions are critical for the early diagnosis and treatment of skin diseases. 
Visual features of skin lesions vary significantly because the images are collected from patients with different lesion colours and morphologies by using dissimilar imaging equipment. 
Recent studies have reported that ensembled convolutional neural networks (CNNs) are practical to classify the images for early diagnosis of skin disorders. 
However, the practical use of these ensembled CNNs is limited as these networks are heavyweight and inadequate for processing contextual information. 
Although lightweight networks (e.g., MobileNetV3 and EfficientNet) were developed to achieve parameters reduction for implementing deep neural networks on mobile devices, insufficient depth of feature representation restricts the performance. 
To address the existing limitations, we develop a new lite and effective neural network, namely HierAttn. 
The HierAttn applies a novel deep supervision strategy to learn the local and global features by using multi-stage and multi-branch attention mechanisms with only one training loss. 
The efficacy of HierAttn was evaluated by using the dermoscopy images dataset ISIC2019 and smartphone photos dataset PAD-UFES-20 (PAD2020). 
The experimental results show that HierAttn achieves the best accuracy and area under the curve (AUC) among the state-of-the-art lightweight networks. 
The code is available at \url{https://github.com/anthonyweidai/HierAttn}.
\end{abstract}


\end{frontmatter}

\section{Introduction}
Skin conditions and disorders are among the most common human diseases to affect millions of people \citep{guy2015prevalence, guy2015vital}. Statistical data shows that around 20\% of Americans are diagnosed with malign cutaneous diseases \citep{esteva2017dermatologist}. Skin cancer, consisting of non-melanoma and melanoma, affected more than 1.5 million new cases globally in 2020. Skin cancer is estimated to be the fifth most common detected cancer in the U.S., with 196,060 new cases reported in 2021 \citep{sung2021global}. The annual cost for treatment of skin cancer is projected to triple from 2011 to 2030 \citep{guy2015vital}. Proactive detection and early diagnosis are critically essential to save patients. For instance, the five-year survival rate for melanoma patients could be 99\% with early-stage diagnosis and treatment, whereas the survival rate is dropped to around 27\% if the conditions are detected in the late stage \citep{sung2021global}. 

Traditional methods for detecting skin disorders include skin cancer screening by self-examination and clinical examination. Among these methods, self-examination is the most common method for the early detection of skin diseases. Around 53\% of patients with melanomas are self-examined before approaching medical experts \citep{aviles2016detects}. Clinical skin examination can provide affirmative screening of skin cancers with a high detection accuracy \citep{loescher2013advances}. However, clinical examinations consume a considerable amount of time for medical professionals to review a large number of dermoscopic images. The long waiting time of weeks or months could delay the treatment and result in unexpected progress of the skin conditions.

The advances in imaging technology make it possible to diagnose skin lesions by analysing optical images of the skin lesions. The imaging modalities in skin examination include reflectance confocal microscopy, total body photography, teledermatology, and dermoscopy. Among all the imaging modalities, dermoscopy is a non-invasive imaging method without reflecting light to examine the skin lesions with up to 10$\times$ magnification \citep{loescher2013advances}. However, manual analysis of dermoscopic images is time-consuming, and the examination results are subjective to healthcare providers. 

Due to the sophisticated features of skin lesion images, it is nontrivial to automatically detect unhealthy skin areas with satisfactory accuracy. Previously, computer-aided approaches were proposed to analyse skin lesion images. Histogram thresholding employs empirical thresholds to isolate lesions from the rest of the skin tissues \citep{celebi2009lesion}. Principle component analysis (PCA) for colour histogram was further utilised in lesion colour space clustering for segmentation \citep{peruch2013simpler}. Gradient vector flow was computed to extract the smoothness and compactness of curve shapes in skin lesions for separating lesions from neighbouring tissues \citep{zhou2011gradient}.  However, the majority of these techniques require heavy human participation and cannot extricate sufficient features from a whole skin lesion image to diagnose skin diseases. 

To achieve an accurate and unbiased diagnosis of skin diseases, deep learning has been applied to extract representative features and provide an end-to-end analysis of medical images. A multi-scale and two-stream convolutional neural network (CNN) was introduced in skin lesions classification \citep{kawahara2016multi}. Moreover, the classification activation feature map was used to capture region of interest information for learning indicative lesion representations \citep{tang2020gp}. Modified ResNet with U-shape CNN architectures were also applied to the classification and segmentation of skin lesions \citep{chen2018multi, xie2020mutual}. In other studies, researchers processed three modalities of patient data, including non-image metadata, clinical and dermoscopy images, in a two-stage feature fusion network to enhance skin lesions classification performance \citep{tang2022fusionm4net}. To learn a different degree of contextual information, Dai et al. introduced an encoder which resizes features in different scales \citep{dai2022ms}. To improve the diagnostic results, researchers assembled several models, such as ResNeXt, NASNet, SENet, DenseNet121 and EfficientNet, to detect skin cancer, and the highest accuracy of these models was reported up to 94.2\% and 92.6\% on the ISIC2018 and ISIC2019 datasets, respectively \citep{adegun2021deep}. However, the majority of reported networks, especially those constructed by combining several models, consumed a large number of computational resources and were highly time-consuming due to the increased complexity of models.

Recently, lightweight algorithms have been reported to promote the use of deep learning in conducting direct skin cancer screening with limited computational resources in clinical computers and mobile devices. One major solution to achieving lightweight CNNs is the depthwise separable convolution (DWSConv), which replaces the standard convolution with depthwise convolution within one channel and pointwise convolution along each channel for the depletion of learnable parameters \citep{howard2017mobilenets}. As for lightweight vision transformer (ViT) networks, sparse attention \citep{pan2022edgevits}, random feature attention \citep{suwanwimolkul2022efficient}, and  low-rank approximation \citep{yang2022lite} were adopted to reduce ViTs' size and computational cost. Most recently, several models leveraged CNN and ViT by applying standard convolution \citep{pan2022edgevits} or DWSConv \citep{mehta2021mobilevit} to downsize tensors before each ViT module. Moreover, knowledge distillation was applied to reduce model parameters \citep{song2022spot}. Although the aforementioned techniques have attained satisfactory results in common object recognition, lightweight deep learning models have not been investigated for skin lesions analysis.

\begin{figure*}[!b]
\centering
\includegraphics[width=\textwidth]{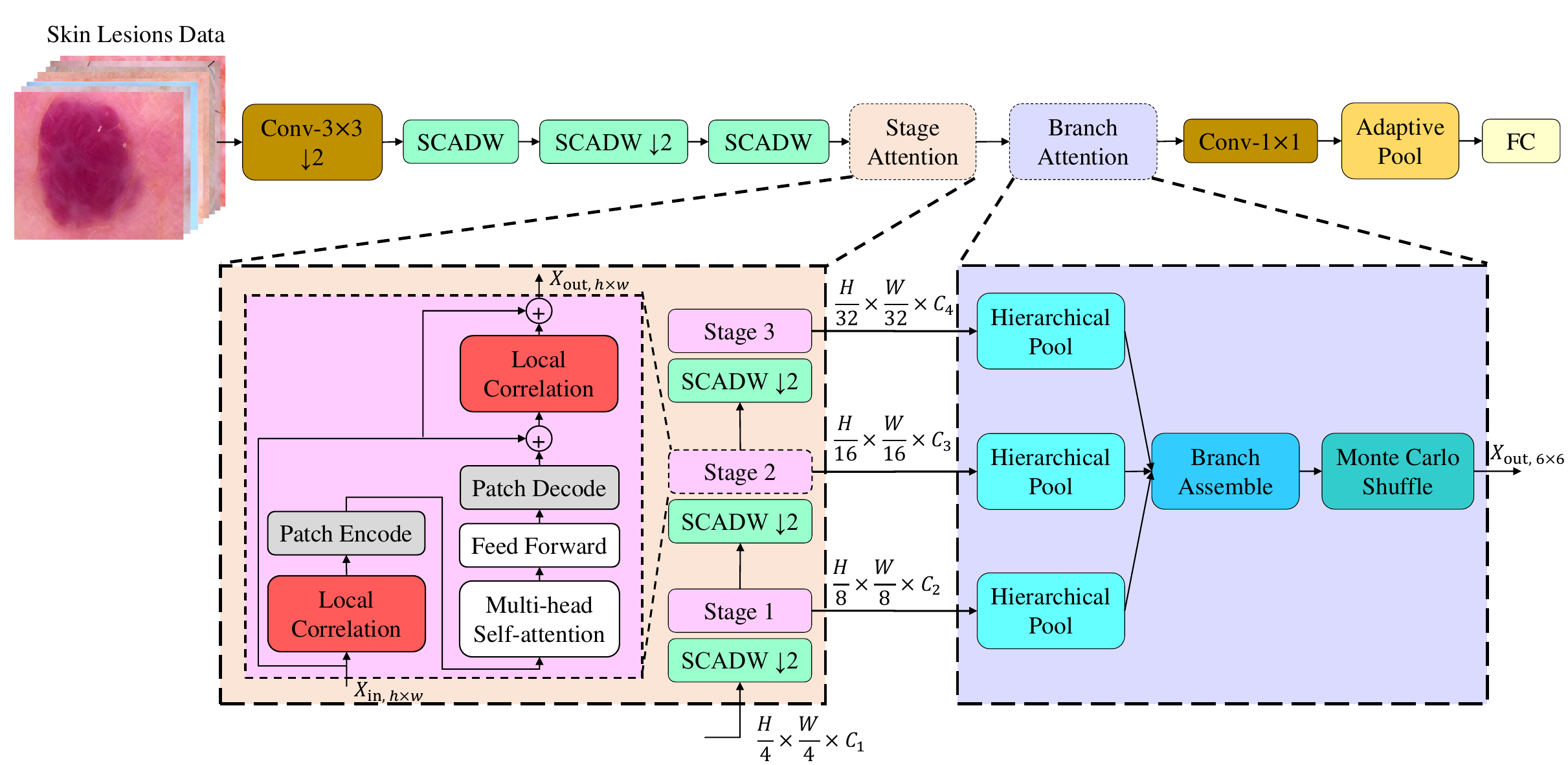}
\caption{{\bf HierAttn architecture.} Conv-$n\times n$ represents a standard convolution, and SCADW refers to a depthwise separable convolution block with the SCAttn module. Down-sampling blocks are marked with ↓2. Stage attention has a SCADW block followed by a CTH block (pink block) that thoroughly aggregates local features and learns contextual representations. Branch attention utilises hierarchical pooling to extract global and local features steadily.}
\label{fig:archi}
\end{figure*}

This article aims to address the challenges in balancing the reliability and accuracy of skin lesion analysis with small memory storage and minimal computational cost. The major novelty and key contributions of this study are:\\

\begin{itemize}
\item We introduce a new lightweight and deeply supervised architecture with diverse attention mechanisms, namely HierAttn, to distinguish multi-class skin lesions. 
HierAttn achieves the top-level performance with a smaller size than other competing mobile models for both the ISIC2019 and PAD2020 datasets.\\
\item We develop a novel deep supervision method, branch attention algorithm, to learn both local and global representations from coarse level to fine level of features with hierarchical pooling and aggregate the extracted hierarchical features by tensors assembling. 
Since the hierarchical pooling and tensor assembling do not introduce additional learnable parameters, high-level features generated by the stage attention block can be extracted without increasing the model size.
Moreover,  branch attention does not introduce additional loss computation, which is computation-friendly compared to the conventional deep supervision method with multiple losses. \\
\item We propose the same channel attention (SCAttn) module based on global averaging pooling without unnecessary modification of channel-wise features. 
The SCAttn design effectively extracts global features and outperforms other attention methods like squeeze and excitation while avoiding the increment of the model parameters to keep a small model size.\\
\item We propose the stage attention block, consisting of a SCAttn block followed by a convolution-transformer hybrid (CTH) block, to thoroughly learn the regional and global high-level feature representations. 
As shown in Fig. \ref{fig:archi}, the HierAttn network involves a novel CTH block by adding a skip connection and stochastic depth to gain optimal performance.\\
\end{itemize}

We review related literature in Sec. \ref{sec:rw} and explain the proposed methodology in Sec. \ref{sec:pm}. We present and discuss the experiment results in Sec. \ref{sec:er} and Sec. \ref{sec:disc}. We summarise the major findings in Sec. \ref{sec:con}.

\section{Related Work}\label{sec:rw}
\subsection{CNNs in skin lesions diagnosis}
The recent advances in machine learning have introduced an increasing number of deep neural networks. The deep learning models, including InceptionV3, VGG, EfficientNet, ResNet, DenseNet, etc., have been applied to classify skin lesions \citep{esteva2017dermatologist, weng2020addressing, gessert2020skin}. Moreover, the models, assembling several networks (e.g., ResNeXt, NASNet, SENet, DenseNet121, and EfficientNet) in several streams or stages, were utilised in skin lesions classification \citep{gessert2020skin, mahbod2020transfer, attique2021two}. The ISIC2019 challenge winner applied an ensemble model accumulating EfficientNet B0 to B6, SENet154, and two ResNeXt to achieve a 92.6\% average classification accuracy; however, the model is comparatively large with more than 100 M parameters, making it impractical for clinical and home use \citep{gessert2020skin}. 

\subsection{Deeply supervised learning}
Adding and widening layers are common methods to improve training metrics by increasing the learning parameters of networks. However, both methods increase the computational complexity and requirements of hardware during training. Deeply supervised learning is frequently used to learn coarse-to-fine features from intermediate branches. Instead of adding layers to the network, deep supervision usually uses auxiliary supervision branches in critical intermediate stages during training \citep{wang2015training}. Multilevel losses are widely used in deep supervision to extract feature information from stages to improve the model’s performance. For instance, GoogleNet has three losses in three branches, separately \citep{szegedy2015going}. Liu et al. applied deep supervision with multiple losses to improve object edges learning \citep{liu2016learning}. However, the features in various stages are fed into simple tensor operations separately. They do not have a direct correlation influence on each other. 

\subsection{Attention mechanisms}
The attention mechanism is a biomimetic cognitive method used in diverse computer vision assignments such as image classification \citep{woo2018cbam, hu2018squeeze, dosovitskiy2020image, hou2021coordinate, mehta2021mobilevit} and image segmentation \citep{mehta2021mobilevit, he2022fully}. An example of the attention network is the SENet, which obtains global representations by global average pooling and channel-wise feature response by squeeze and excitation \citep{hu2018squeeze}. The convolution block attention module (CBAM) improves the SENet by using a larger kernel size to encode spatial information \citep{woo2018cbam}. Coordinated attention further advances the global average pooling by encoding channel relationships and long-range dependencies via average pooling along different axes \citep{hou2021coordinate}. However, they introduce more learnable parameters and consume more computational resources than the SENet. 

\subsection{Vision transformer}
To improve the computational efficiency and meet the scalability requirement, self-attention mechanisms, particularly transformers, are introduced into computer vision from natural language processing. With self-attention, a ViT replaces the traditional convolution method with a transformer encoder \citep{dosovitskiy2020image}. Because the input images are directly split into patches and embedded, the ViT models are still very large, with more than 100 M parameters \citep{dosovitskiy2020image}. The standard transformer has been applied to process sequences of image patches to learn the inter-patch representations. However, the original transformer method ignores the inductive biases (e.g., translation equivariance and locality) inherent to CNNs, which leads to poor performance when training with insufficient data \citep{dosovitskiy2020image}.  Recently, ViT was utilised in skin lesions evaluation by appending a ViT branch to CNN-based U-shape architecture for improving long-range dependencies and contextual information extraction, but an additional branch notably increased the model weight \citep{wu2022fat}.

\subsection{Lightweight networks}
To reduce computational cost and model weight, researchers proposed several well-accepted methods such as knowledge distillation, DWSConv, and efficient self-attention design. Although knowledge distillation can effectively reduce learnable parameters, it bares the huge computational cost of the training process. Moreover, complex teacher and student network structures and parameters updated between the two networks should be carefully designed \citep{reiss2021every}. In addition, based on DWSConv method, MobileNetV2, MobileNetV3, EfficientNet, MnasNet, and ShuffleNet were constructed and obtained relatively satisfactory performance \citep{sandler2018mobilenetv2, zhang2018shufflenet, tan2019efficientnet, tan2019mnasnet, howard2019searching}. To reduce the latency of splitting images, computer scientists from Apple proposed MobileViT to tackle the loss of inductive biases by taking convolution and transformer to form a hybrid block \citep{mehta2021mobilevit}. However, it can not entirely learn representations because the number of its layers is deficient. Moreover, to the best of our knowledge, the aforementioned networks have not been applied to skin lesion analysis, and their performance remains unknown.

\section{Proposed Methodology}\label{sec:pm}
\subsection{Skin lesion dataset}
\begin{figure}[!h]
\centering
\includegraphics[width=\columnwidth]{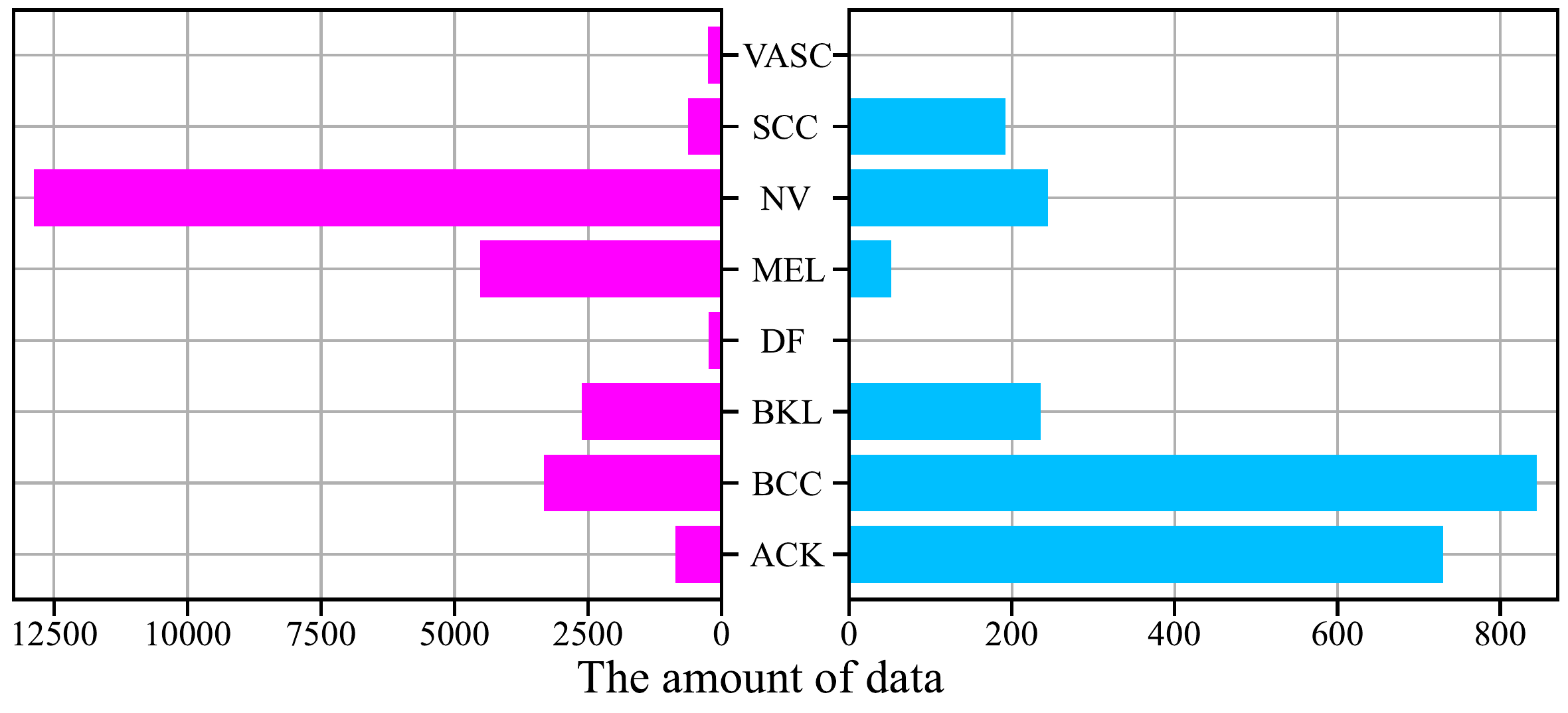}
\caption{{\bf Data distribution} on ISIC2019 (left image) and PAD2020 (right image)}
\label{fig:data-dis}
\end{figure}

Two publicly available skin lesions datasets, ISIC2019 \citep{tschandl2018ham10000, adegun2021deep, combalia2019bcn20000} and PAD2020 \citep{pacheco2020pad}, are used in this research for development and evaluation of the proposed methods. Dermoscopy and smartphones are two standard methods to capture skin lesions images. Thus, the two datasets effectively represent current image data for the classification of skin lesions. The data distribution of the two datasets is shown in Fig. \ref{fig:data-dis}.

ISIC2019 dataset consists of 25,331 dermoscopy images with eight categories: actinic keratosis (ACK), basal cell carcinoma (BCC), benign keratosis (BKL), dermatofibroma (DF), melanoma (MEL), melanocytic nevus (NV), squamous cell carcinoma (SCC), vascular lesion (VASC). Three of them (ACK, BCC and SCC) belong to the non-melanoma skin cancer. Melanoma is the most severe skin cancer that is caused by uncontrolled growing cells that can produce pigment. The vascular lesions (VASC) can be either benign or malign and requires close monitoring over a period of time. The remaining categories of skin lesions are benign conditions. The PAD2020 dataset includes 2,298 skin lesion images with six classes: ACK, BCC, BKL, MEL, NV, and SCC, collected by using smartphones. Compared to ISIC2019, PAD2020 has two fewer classes, DF and VASC, because of lacking photos of skin lesions. 

\subsection{Pre-processing}
\subsubsection{Image pre-processing}
\begin{figure}[h]
     \centering
     \begin{subfigure}[b]{0.24\columnwidth}
         \centering
         \includegraphics[width=\textwidth]{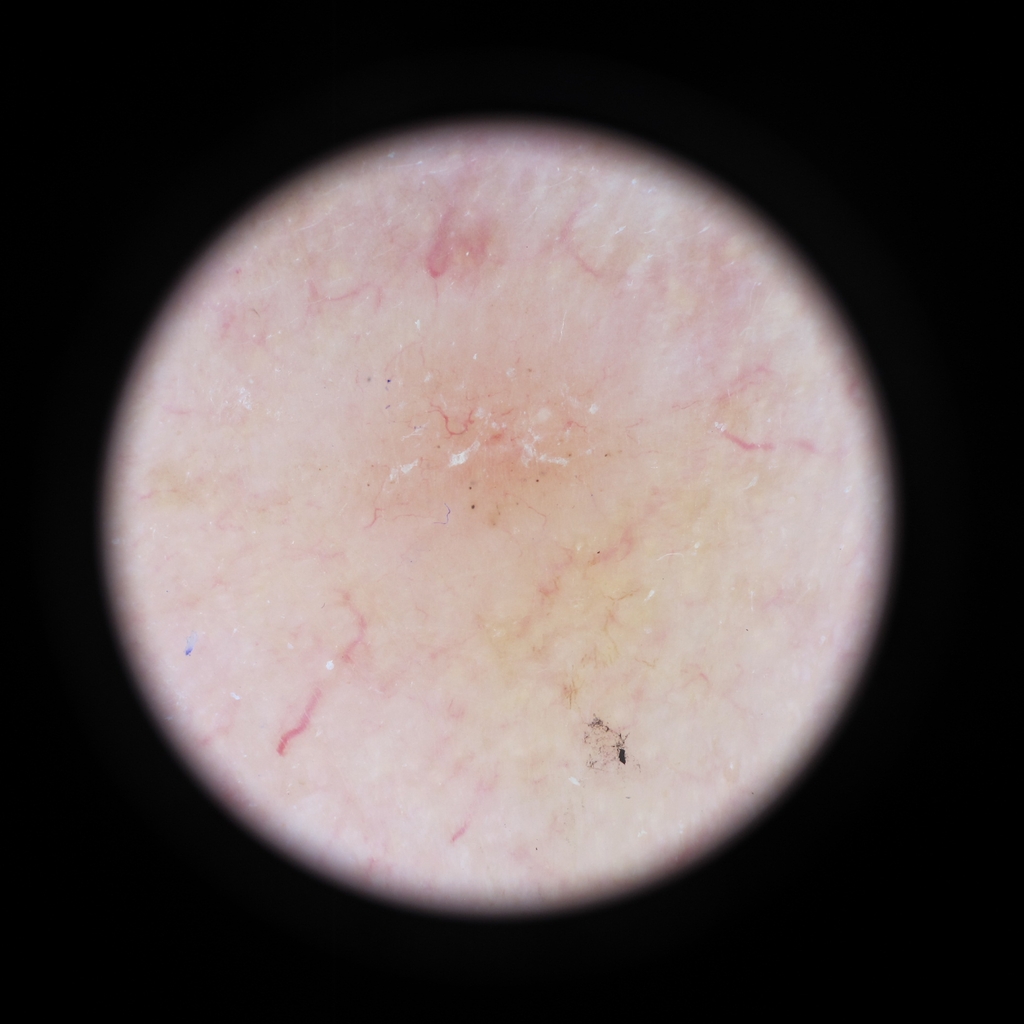}
         \caption{}
         \label{fig:ce1}
     \end{subfigure}
     \hfill
     \begin{subfigure}[b]{0.24\columnwidth}
         \centering
         \includegraphics[width=\textwidth]{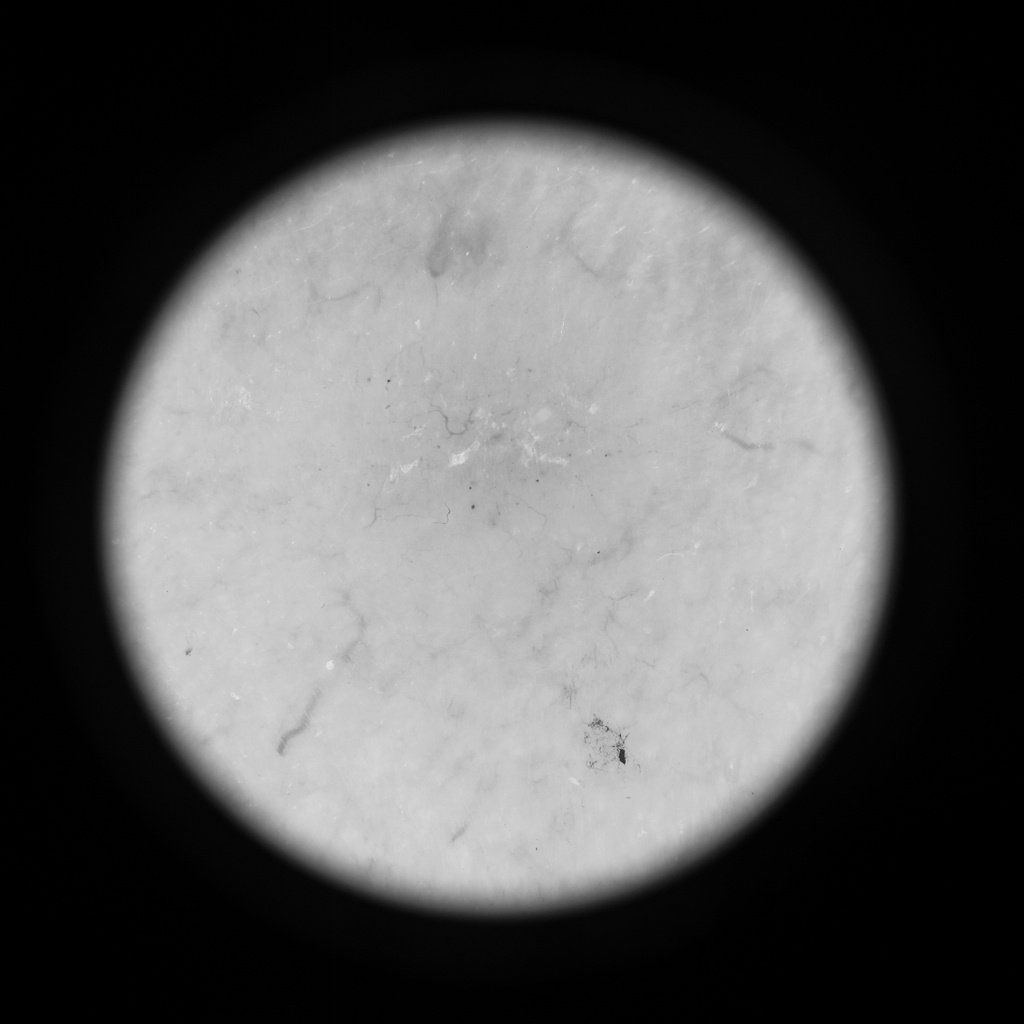}
         \caption{}
         \label{fig:ce2}
     \end{subfigure}
     \hfill
     \begin{subfigure}[b]{0.24\columnwidth}
         \centering
         \includegraphics[width=\textwidth]{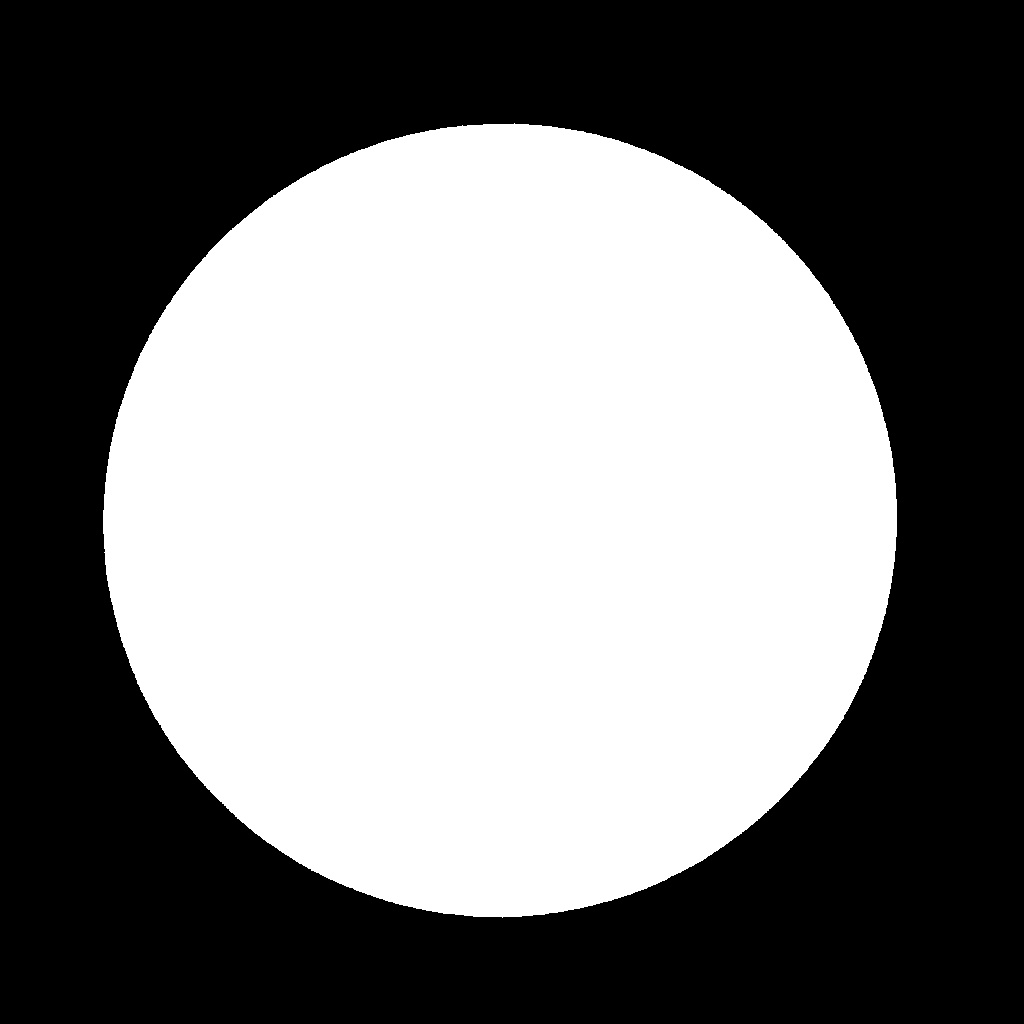}
         \caption{}
         \label{fig:ce3}
     \end{subfigure}
     \hfill
     \begin{subfigure}[b]{0.24\columnwidth}
         \centering
         \includegraphics[width=\textwidth]{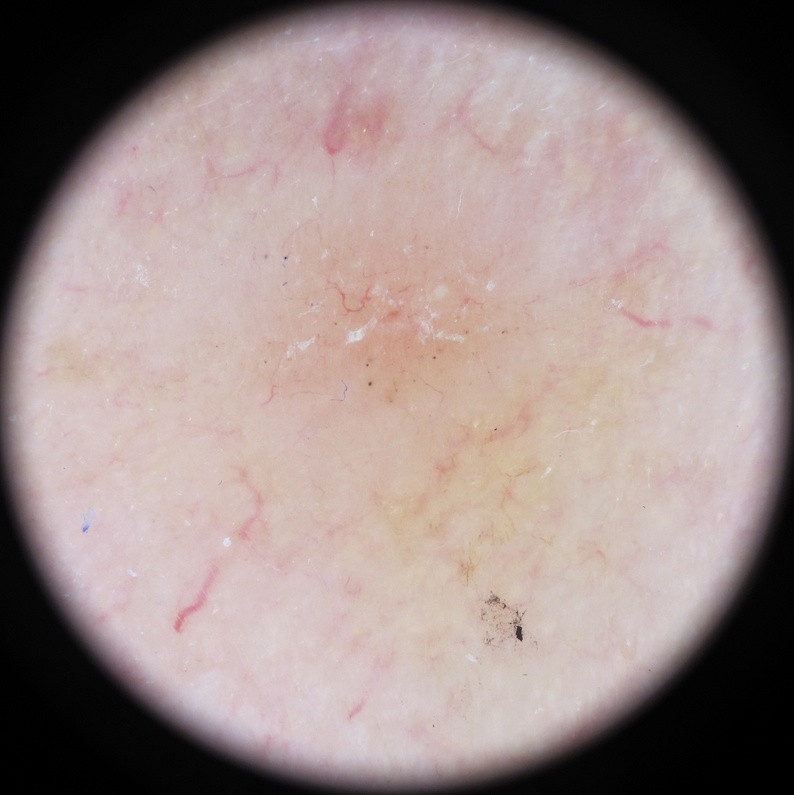}
         \caption{}
         \label{fig:ce4}
     \end{subfigure}
        \caption{{\bf The progress in cropping image} (a) original image, (b) greyed image, (c) binarised image, and (d) cropped image.}
        \label{fig:ce}
\end{figure}

The data among different classes from the ISIC2019 dataset has a large black area [see Fig. \ref{fig:ce} (a)], which damages the evaluation performance of deep learning models. Thus, an adaptive cropping method is taken to identify and crop these images. The original image is first turned into greyscale [Fig. \ref{fig:ce} (b)] and then binarised an adaptive threshold ranging from 50 to 255 [Fig. \ref{fig:ce} (c)]. After that, the contour of the binarised image is detected to confirm the circle location. When the value of the circle area divided by the whole image area is between 0.01 and 0.9, the region enclosed by the circle is cropped and saved [Fig. \ref{fig:ce} (d)].

\subsubsection{Data balance}
In this study, an imbalance ratio is defined to assess the imbalance problem quantitatively. The imbalance ratio is a fraction of the number of images of the majority class over the minority class. The imbalance ratio for ISIC2019 and PAD2020 are 53.9 and 16.3, respectively. Such relatively high imbalance ratios can remarkably reduce the performance of training, according to previous research \citep{buda2018systematic}. The data imbalance could lead to a low validation result for the minor class, though the averaged validation metrics over all classes could be high. Data balance by oversampling or undersampling for each class is a practical technique to handle the dataset with a huge imbalance ratio. Oversampling and undersampling are simultaneously utilised to balance ISIC2019 and PAD2020 data. After sampling, 2500 images and 500 images are collected for each class in the ISIC2019 and PAD2020 datasets, respectively. Thus, the amount of data after balancing totals 20000 and 3000 on ISIC2019 and PAD2020, respectively, is close to the total amount of unbalanced data.

Oversampling uses horizontal flips, random crops, Gaussian blur, linear contrast, random translation, rotation, and shear on a small scale to generate new images from the old images. As for undersampling, the random selection of a fixed number of images tends to have a class-overlapping problem. Thus, we apply an adaptive data analysis method called instance hardness (IH) to alleviate this adverse effect. 
Instance hardness is defined as:

\begin{equation}
\text{IH}_\mathcal{L}\left(\left<x_i,\ y_i\right>\right)=1-\frac{1}{\left|\mathcal{L}\right|}\Sigma_{j=1}^{\left|\mathcal{L}\right|}p(y_i|x_i,g_j(t,\alpha))
\end{equation}
where $\mathcal{L}$ is a prior with non-zero probability while treating all other learning algorithms as having zero probability, $g$ is a machining learning algorithm trained on $t$ with the hyperparameter $\alpha$, and $y_i$  is the label for data $x_i$.

Outliers and mislabelled data are expected to have high IH. Thus, IH analysis was applied in undersampling to remove those data with a high IH. This research uses the random forest as the machine learning method g, referenced from the imbalanced-learn study \citep{smith2014instance, lemaitre2017imbalanced}. 

\subsection{HierAttn architecture}
The paper introduces an efficient and lite Convolution-Transformer model [see Fig. \ref{fig:archi}], HierAttn, for skin lesions diagnosis. The main idea of this model is to hierarchically learn the local and global representations by utilising various attention mechanisms. Branch attention allows the network to extract various levels of features from different layers. Moreover, the SCAttn module in our new HierAttn network leverages the DWSConv by supplementing global information to improve performance. The three novel mechanisms (i.e., SCAttn, stage attention, and branch attention) consume miniature parameters or no learnable parameter, which results in the tiny size of HierAttn. Besides, the stage attention utilises an effective self-attention hybridising convolution to keep inductive biases and reduce the latency, inspired by the MobileViT architecture \citep{mehta2021mobilevit}. The attention modules are described in the following sections. Detailed structures of HierAttn are shown in Table \ref{tab:hierattn}.

\subsubsection{Branch attention}
\begin{figure}[!h]
\centering
\includegraphics[width=\columnwidth]{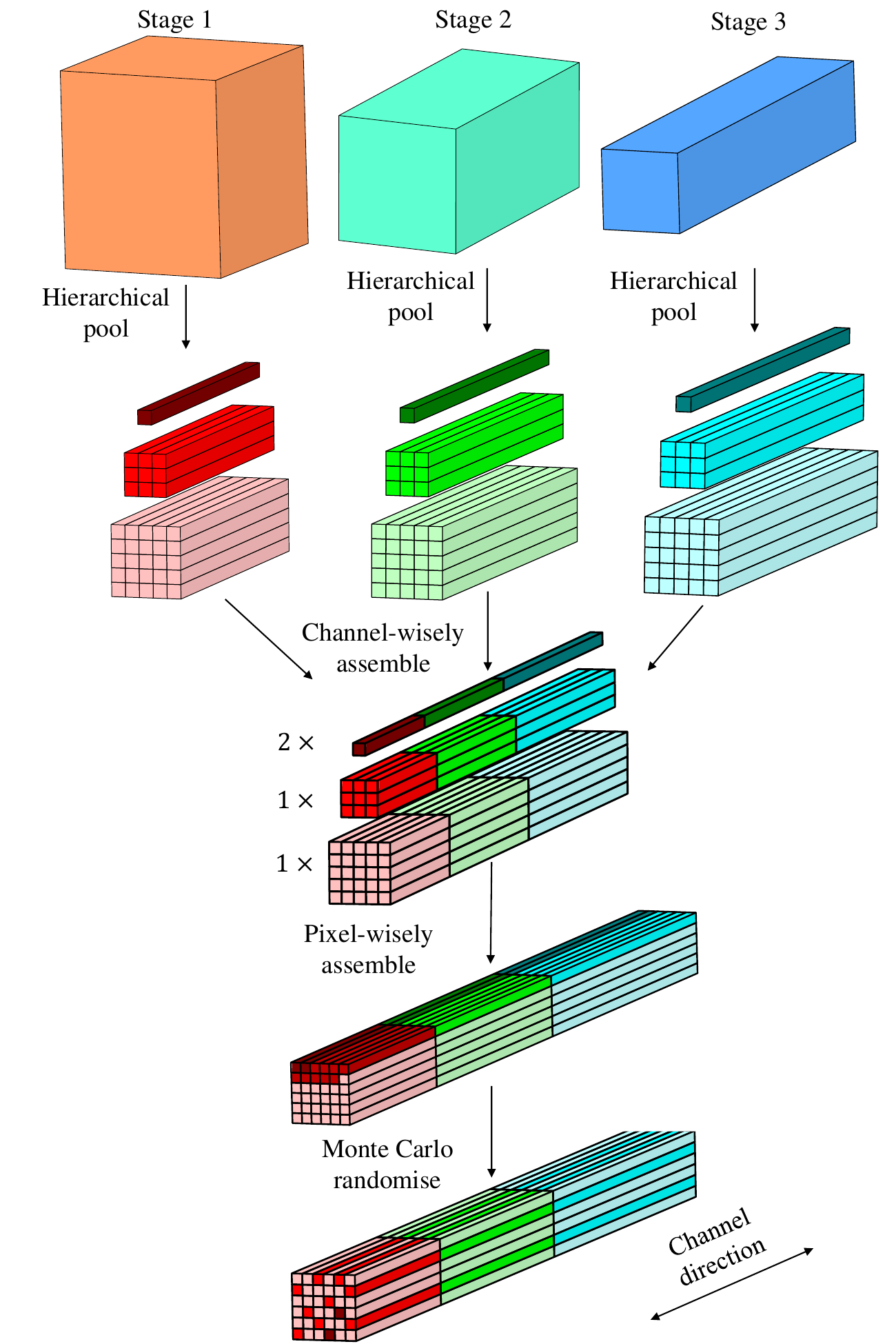}
\caption{\bf Branch attention by hierarchical pooling, assembling and randomising tensors from different branches.}
\label{fig:branchattn}
\end{figure}

Zhang et al. employed a contrastive loss to improve the feature interactions during training among different intermediate stages \citep{zhang2022contrastive}. Instead, we propose a novel deep supervision method, brach attention (i.e., lower right in Fig. \ref{fig:archi}), utilising hierarchical pooling to downsize tensors from different stages, tensors assembling to improve the interactions of the downsized features and learn the hierarchy of features from the assembled tensors. Moreover, by keeping the different sizes of pooling results, hierarchical pooling learns the local representations of tensors, generating large tensor size ($C\times 5 \times 5$), and attains tensors’ global representations, generating small tensor size ($C\times 1\times 1$). Additionally, the medium output tensor with size $C\times 3\times 3$ is also designed as a buffer layer to keep both local and global features. The pooled tensors from different branches are then channel-wisely assembled. After that, the ensembled tensors are pixel-wisely assembled. At the end of branch attention, the pixel-wisely assembled tensors are pixel-wisely shuffled by utilising the Monte Carlo method. The branch attention is applied as a particular learning stage after each stage attention block (critical learning stage). Most importantly, by using such a practical design, branch attention does not introduce additional loss, which cost less computational cost than the conventional deep supervision method. The branch attention progress is illustrated in Fig. \ref{fig:branchattn}.

\subsubsection{Same channel attention} 
The block modifies the DWSConv with the SCAttn technique and results in a new block structure called SCADW block (i.e., green blocks in  Fig. \ref{fig:archi}). Squeeze and excitation attention (SEAttn) blocks were proposed to enhance the expressive power of the learned features after depthwise convolution in the MobileNetV3 block \citep{howard2019searching}. However, the pointwise convolution already extracts the channel-wise information, which suggests that the excitation of SEAttn is likely a redundant operation. Therefore, the SCAttn is introduced after depthwise convolution by global average pooling while maintaining the same number of learnable parameters as the DWSConv block. The SCAttn block is illustrated in Fig. \ref{fig:scattn}.

\subsubsection{Stage attention}
\begin{figure}[!h]
     \centering
     \begin{subfigure}[b]{0.48\columnwidth}
         \centering
         \includegraphics[width=\textwidth]{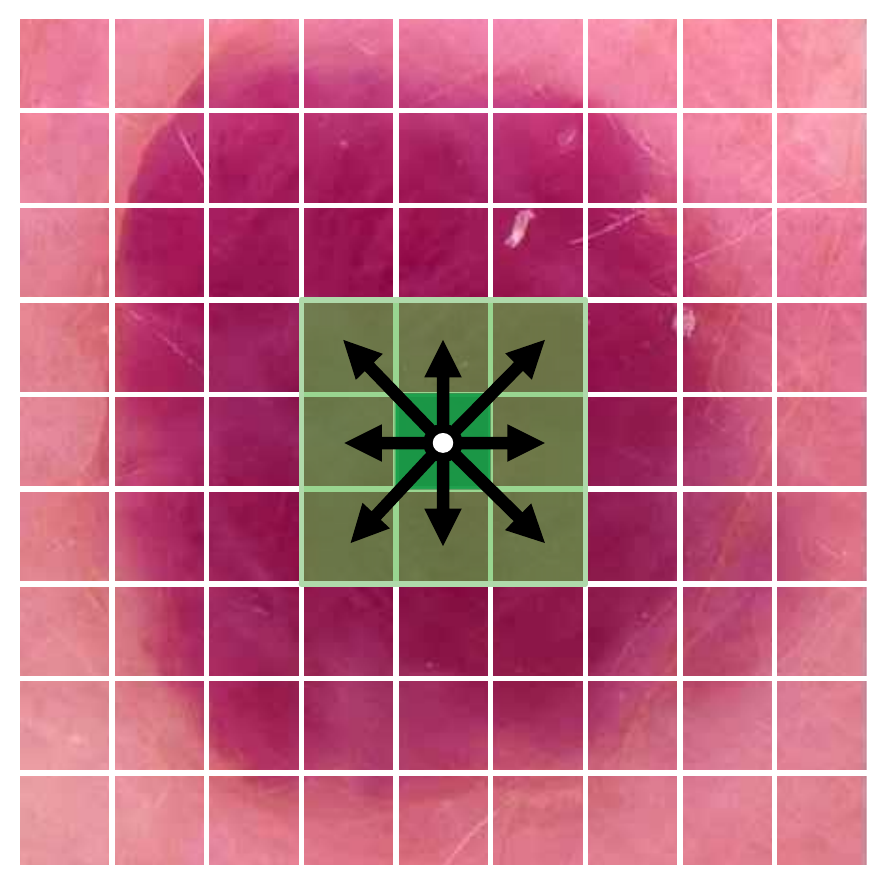}
         \caption{}
         \label{fig:sa1}
     \end{subfigure}
     \hfill
     \begin{subfigure}[b]{0.48\columnwidth}
         \centering
         \includegraphics[width=\textwidth]{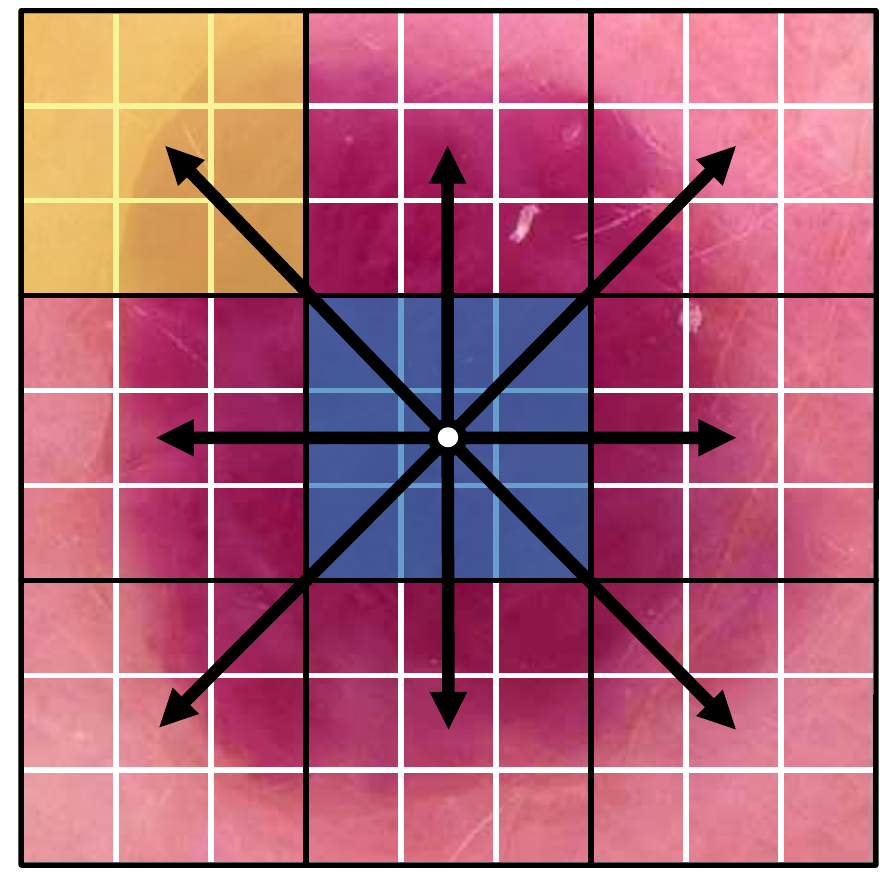}
         \caption{}
         \label{fig:sa2}
     \end{subfigure}
        \caption{{\bf Feature interactions in stage attention.} (a) Local correlation extract local features and (b) Multi-head self-attention further learns global information.}
        \label{fig:sa}
\end{figure}

Each stage attention module (i.e., lower left in Fig. \ref{fig:archi}) has a SCADW block with a stride of 2 followed by a convolution-transformer hybrid block. We apply the convolutions and transformers simultaneously to learn the local and global representations of an input skin lesion image with fewer parameters. The convolution-transformer hybrid (CTH) block (i.e., pink blocks in Fig. \ref{fig:archi}) utilised consecutive modules to process feature maps before and after encoding. The CTH block consists of a multi-head self-attention (MHSA) followed by a multilayer perceptron feed forward (FFN) layer, which is regarded as vision trasnformer (ViT). ViT learns long-range spatial dependence among encoded patches, which is shown in Fig. \ref{fig:sa} (a). To integrate local features, local correlation through standard convolution is applied before and after the ViT module in the CTH block, which is illustrated by Fig. \ref{fig:sa} (b). Furthermore, we also add a skip connection to link the input and output of the CTH block. Stage attention module thoroughly rearranges feature maps by downsizing feature maps, and the CTH block further extracts the processed features. Thus, each stage attention module is regarded as a {\bf critical learning stage} in HierAttn. The CTH block bottleneck can be formulated as:

\begin{equation}
 \begin{gathered}
X = \text{LocalCorr}(X_\text{in}) + X_\text{in} \\
Y = \text{MHSA}(\text{Encode}(X)) \\
Z = \text{Decode}(\text{LayerNorm}(\text{FFN}(Y))) \\
X_\text{out} = \text{LocalCorr}(Z + X_\text{in}) + X_\text{in}
 \end{gathered}
\end{equation}

\subsubsection{Small-scale attention modules}
SCAttn leverages SEAttn by keeping single-pixel attention through global averaging pooling and avoiding the modification of tensors along channel direction. It contributes to utilising an attention mechanism to extract global features after depthwise convolution without introducing any learnable parameters. Moreover, branch attention does not increase model size because hierarchical pooling and tensor assembling in branch attention are parameter-free. In addition, transformer operations lose spatial bias, increasing computational costs with a wider and deeper design to learn visual representations \citep{mehta2021mobilevit}. Therefore, ViT-Base, ViT-Large, and ViT-Huge models use the number of transformer blocks L = 12, 24, 32 and the number of embedded dimensions d = 768, 1024, 1280, respectively \citep{dosovitskiy2020image}. However, our new HierAttn only requires L = {2, 4, 3} and d = {96, 120, 144}. The number of parameters for the new HierAttn and other state-of-the-art (SOTA) networks is summarised in Table \ref{tab:top1models}.

\begin{figure}[h]
\centering
\includegraphics[width=\columnwidth]{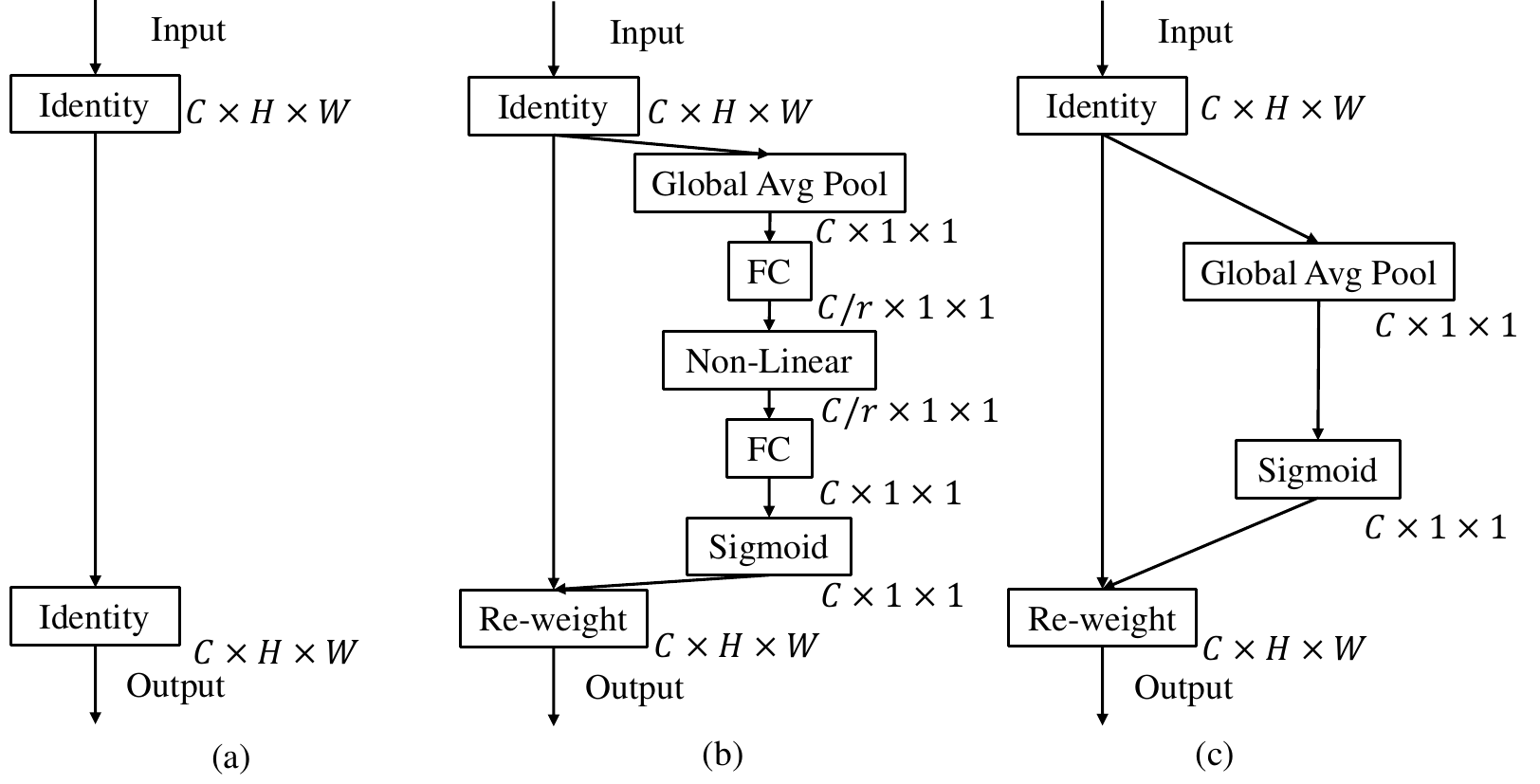} 
\caption{Schematic comparison of the original block (without attention mechanism) (a), the SEAttn block (b), and {\bf the proposed SCAttn block} (c).}
\label{fig:scattn}
\end{figure}

%
%
%

\begin{table*}[!h]
\caption{\label{tab:hierattn} {\bf Detailed structures of the proposed HierAttn architecture.} Here, $d$ means dimensionality of the input to the transformer layer in the CTH block. In the CTH block, kernel size is set as three and patch height and width are set as two.}
\centering
\begin{tabular}{|M{2.2cm}|M{2.2cm}|M{2cm}|M{2cm}|M{1.8cm}|M{1.8cm}|} \hline
\multicolumn{2}{|c|}{\multirow{2}{*}{\bf Layer}} & \multirow{2}{*}{\bf Output size} & \multirow{2}{*}{\bf Repeat} & \multicolumn{2}{c|}{\bf Output Channels}\\
\cline{5-6} \multicolumn{2}{|c|}{} & & & XS &S\\ \hline

\multicolumn{2}{|c|}{Image} & $256\times 256$ & & & \\ \hline

\multicolumn{2}{|c|}{Conv$3\times 3$, $\downarrow 2$} & \multirow{2}{*}{$128\times 128$} & 1 & 16 & 16\\ 
\multicolumn{2}{|c|}{SCADW} & & 1 & 16 & 32\\ \hline

\multicolumn{2}{|c|}{Conv$3\times 3$, $\downarrow 2$} & \multirow{2}{*}{$64\times 64$} & 1 & 24 & 48\\ 
\multicolumn{2}{|c|}{SCADW} & & 2 & 24 & 48\\ \hline

\multirow{2}{*}{Stage1} & Conv$3\times 3$, $\downarrow 2$ & \multirow{2}{*}{$32\times 32$} & 1 & 48 & 64\\ 
& CTH block & & 1 & 48($d=64$) & 64($d=96$)\\ \hline

\multirow{2}{*}{Stage1} & Conv$3\times 3$, $\downarrow 2$ & \multirow{2}{*}{$16\times 16$} & 1 & 64 & 80\\ 
& CTH block & & 1 & 64($d=80$) & 80($d=120$)\\ \hline

\multirow{2}{*}{Stage1} & Conv$3\times 3$, $\downarrow 2$ & \multirow{2}{*}{$8\times 8$} & 1 & 80 & 96\\ 
& CTH block & & 1 & 80($d=96$) & 96($d=144$)\\ \hline

\multicolumn{2}{|c|}{Branch attention} & \multirow{2}{*}{$8\times 8$} & 1 & 192 & 240\\ 
\multicolumn{2}{|c|}{Conv$1\times 1$} & & 1 & 768 & 960\\ \hline

\multicolumn{2}{|c|}{Pooling} & \multirow{2}{*}{$1\times 1$} & \multirow{2}{*}{1} & \multirow{2}{*}{8 or 6} & \multirow{2}{*}{8 or 6}\\ 
\multicolumn{2}{|c|}{Linear} & & & & \\ \hline

\multicolumn{2}{|c|}{{\bf \# Parameters}} & & & 1.08 M & 2.14 M\\ \hline
\end{tabular}
\end{table*}

\begin{table*}[!h]
\caption{\label{tab:top1models} {\bf \# Parameters and accuracy of different models.}}
\centering
\begin{tabular}{|M{6cm}|M{2.5cm}|M{2.2cm}|M{2.2cm}|} \hline
\multirow{2}{*}{\bf Model} & \multirow{2}{*}{\bf \# Parameters/M} & \multicolumn{2}{c|}{\bf Accuracy/\%}\\ 
\cline{3-4} & & IHISIC20000 $\Uparrow$ & IHPAD3000\\ \hline
MobileNetV2\citep{sandler2018mobilenetv2} & 2.23 & 93.45 & 87.44 \\ \hline
MobileViT\_s\citep{mehta2021mobilevit}  & 4.94  & 94.72 & 88.22\\ \hline
MobileNetV3\_Large\citep{howard2019searching} & 4.21 & 94.77 & 88.78\\ \hline
ShuffleNetV2\_1$\times$\citep{zhang2018shufflenet} & 2.28 & 95.23 & 87.89\\ \hline
MnasNet1.0\citep{tan2019mnasnet} & 3.11 & 95.45 & 90.33\\ \hline
EfficientNet\_b0\citep{tan2019efficientnet} & 4.02 & 95.48 & 90.22\\ \hline
{\bf HierAttn\_xs (ours)} & {\bf 1.08} & 96.15 & 90.11\\ \hline
{\bf HierAttn\_s (ours)} & 2.14 & {\bf 96.70} & {\bf 91.22}\\ \hline
\end{tabular}
\end{table*}

\section{Experimental Results}\label{sec:er}
\begin{figure*}[!b]
\centering
\includegraphics[scale=.6]{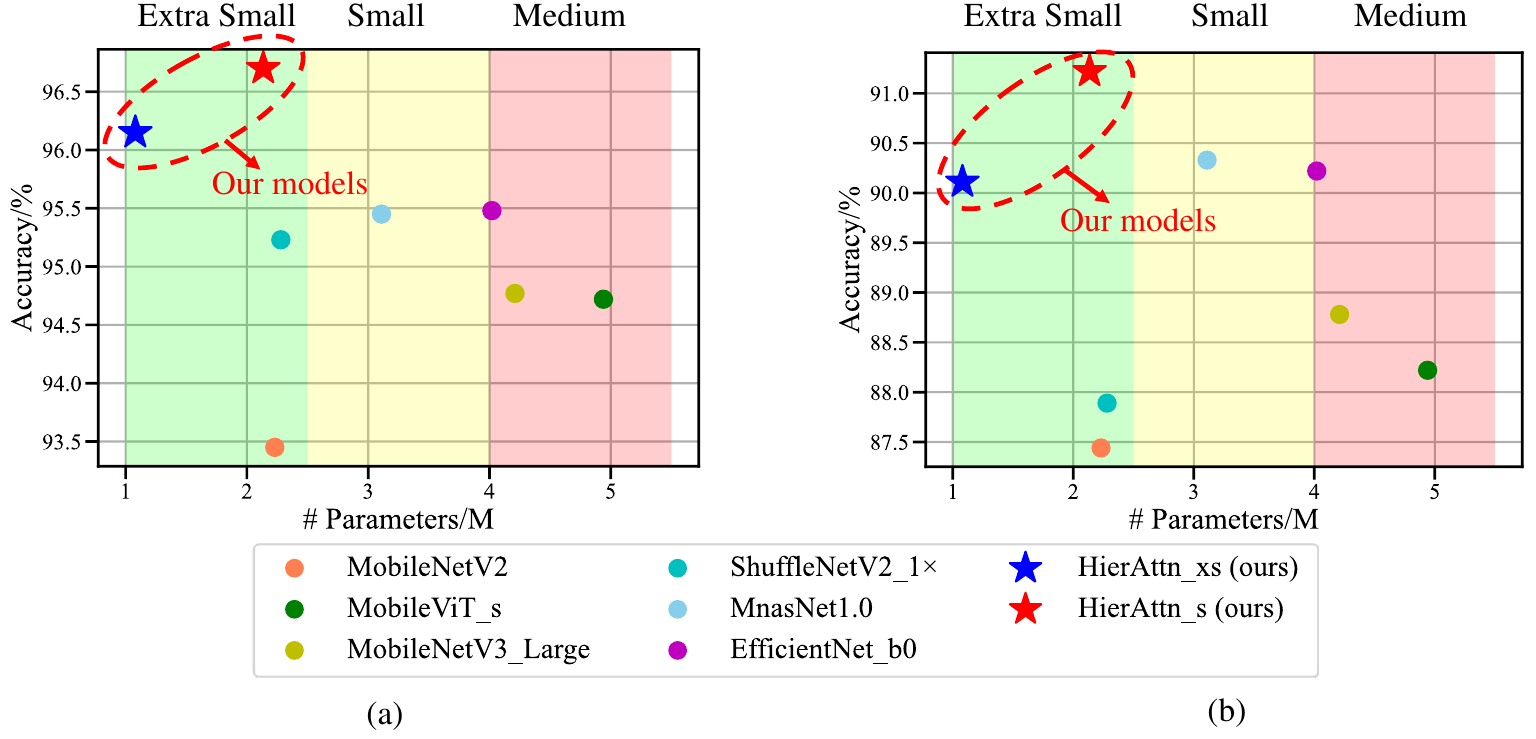}
\caption{{\bf HierAttn vs. SOTA models} on (a) ISIC2019 and (b) PAD2020 validation set.}
\label{fig:top1models}
\end{figure*}

In this section, we first evaluate HierAttn performance on the ISIC2019 and PAD2020 datasets in Sec. \ref{sec:eric}. HierAttn delivers significantly better performance than the SOTA mobile networks ($<$6 M), which can be seen in Table \ref{tab:top1models} and  Fig. \ref{fig:top1models}. In Sec. \ref{sec:eras}, we conducted ablation studies for attention mechanisms in DWSConv. 

\subsection{Evaluation metrics}\label{sec:evametrics}
The ISIC community recommends accuracy, ROC, and AUC to be the evaluation metrics for skin lesions classification \citep{cassidy2022analysis}. In our experiments, top-1 accuracy (abbrev. accuracy) is adopted to evaluate the performance of skin lesions classification (e.g., the proportion of correct values in inference results). Moreover, the receiver operating characteristic (ROC) curve and its area under the curve (AUC) are applied to demonstrate the more general performance of each model (e.g., the model's robustness and scale invariance). Furthermore, each model's number of parameters (\# parameters) is applied to measure its weight.

\begin{figure*}[!b]
     \centering
     \begin{subfigure}[b]{\columnwidth}
         \centering
         \includegraphics[scale=.3]{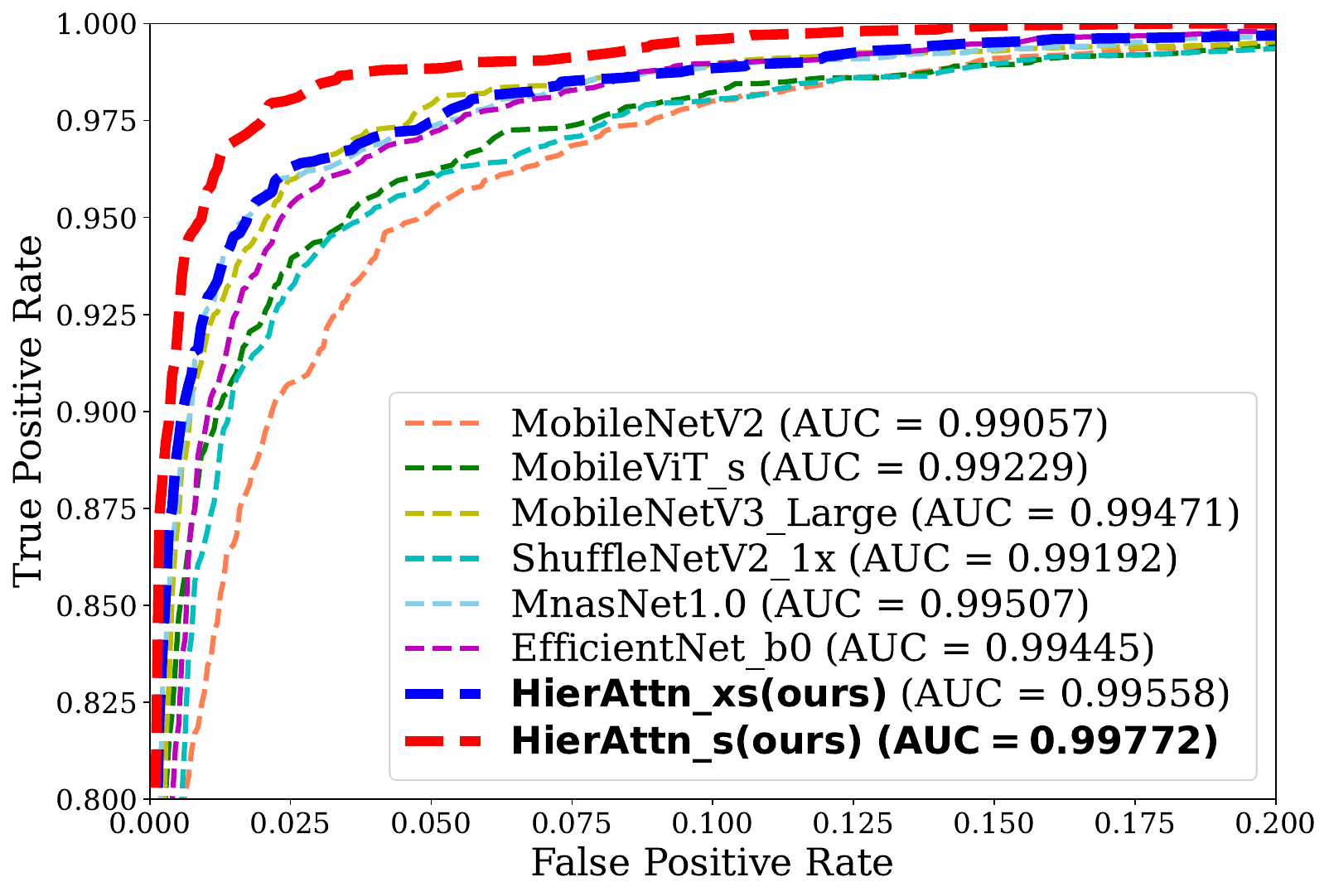}
         \caption{}
         \label{fig:roc1}
     \end{subfigure}
     \hfill
     \begin{subfigure}[b]{\columnwidth}
         \centering
         \includegraphics[scale=.3]{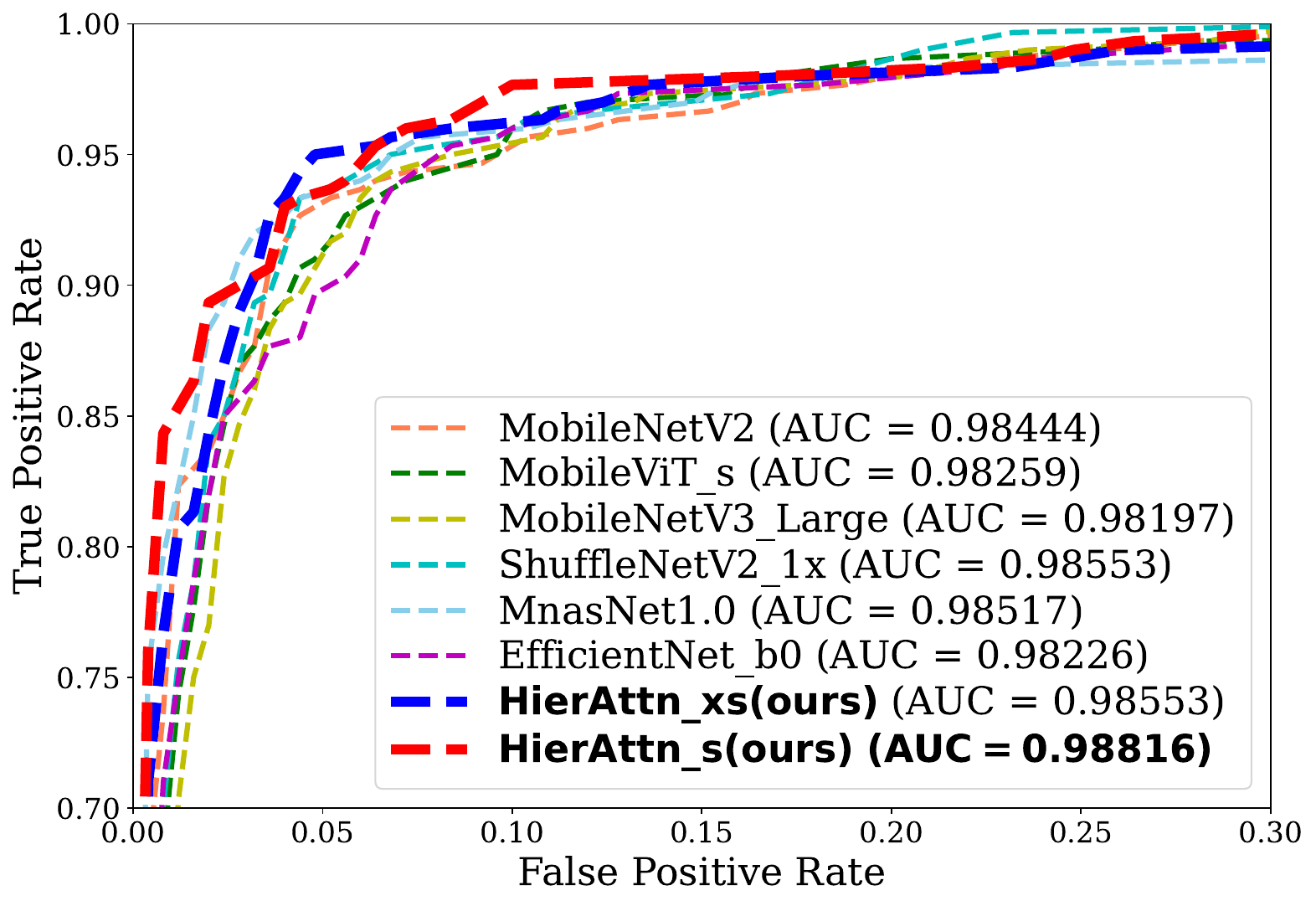}
         \caption{}
         \label{fig:roc2}
     \end{subfigure}
        \caption{{\bf ROC curves of different classification methods on skin lesions dataset} (a) ISIC2019 validation set and (b) PAD2020 validation set.}
        \label{fig:roc}
\end{figure*}

\subsection{Image classification on the skin lesions dataset}\label{sec:eric}
\subsubsection{Implementation details}\label{sec:impdetails}
The HierAttn network and other SOTA lite models are trained and validated for 500 epochs on one RTX A4000 with a batch size of 64 images using AdamW optimiser \citep{loshchilov2018decoupled} with 10-fold cross-validation and cross-entropy loss. The learning rate is ceased from 0.002 to 0.0002 during the first 30 epochs and then increased to 0.0002 utilising the cosine scheduler \citep{loshchilov2017sgdr}. L2 weight decay of 0.01 is adopted. Moreover, knowledge transfer is applied to reduce the training time and improve model performance. All transferred models for training in ISIC2019 were trained in ImageNet1K. The well-tuned models from ISIC2019 were then tuned in PAD2020. And the parameters of other layers of HierAttn after branch attention are randomly initialised. Additionally, the transfer learning warm-up is applied to alleviate the negative influence of untransferred layers on transferred layers. On the first training 30 epochs, all transferred layers are frozen. After that, gradient calculation is required for all layers with learnable parameters. Furthermore, each model with 8 classes on ISIC2019 or 6 classes on PAD2020 has a similar number of parameters when two decimal places are kept. Thus, the number of parameters of each model is computed with eight classes to simplify the discussion. Meanwhile, mobile models with (1, 2.5) M, (2.5, 4) M, (4, 5.5) M parameters are regarded as "extra small", "small", and "medium", respectively.

\subsubsection{Accuracy}
Fig. \ref{fig:top1models} compares HierAttn with six other lightweight networks that are also trained on the ISIC2019 and PAD2020 datasets. Detailed values are illustrated in Table \ref{tab:top1models}. Fig. \ref{fig:top1models} also demonstrates that HierAttn networks fall in the upper left region, which means they outperform other mobile architectures with relatively small sizes. For instance, with about 1.08 million parameters, HierAttn\_s outperforms MobileNetV2 by 3.25\%, MobileNetV3\_large by 1.93\%, ShuffleNetv2\_1x by 1.47\%, MnasNet1.0 by 1.25\%, and EfficientNet-b0 by 0.55\% on the ISIC2019 validation set. Furthermore, HierAttn\_s also outperforms MobileViT\_s, which also contains convolution-transformer hybrid blocks, by 1.98\% and 3.00\% on ISIC2019 and PAD2020 datasets, respectively. Besides, HierAttn attains more than 90\% accuracy on both the dermoscopy and the smartphone skin lesions image datasets, which demonstrates that HierAttn is versatile in visually diagnosing skin lesion images from various domains.

\subsubsection{ROC and AUC}
Fig. \ref{fig:roc} illustrates that all models on experiments have more than 0.99 AUC on the ISIC2019 and 0.98 AUC on the PAD2020 validation sets, suggesting that they possess sufficient analytical capacities to conduct multi-classes lesions classification tasks. Furthermore, the ROCs of HierAttn\_s and HierAttn\_xs are the first and second closest to the point (0, 1) on both ISIC2019 and PAD2020 validation sets. Moreover, HierAttn\_s and HierAttn\_xs have the first and the second largest AUC on both datasets among all models, for instance, 0.99772 and 0.99558 on the ISIC2019 validation set, respectively. Thus, the ROCs and AUCs show that HierAttn is the most reliable and superlative model to detect skin lesions among current conventional and advanced mobile models. Meanwhile, the same model’s AUC of the PAD2020 validation set is lower than the ISIC2019 validation set, which means these models perform better on the ISIC2019 validation set (in conformity with Table \ref{tab:top1models}).

\subsection{Ablation studies}\label{sec:eras}
\subsubsection{Implementation details}
In ablation studies, the model used is HierAttn\_s, and the dataset used is ISIC2019, if not mentioned. Other parameters are the same as Section \ref{sec:impdetails}.

\subsubsection{Data balance methods}
\begin{figure}[!h]
\centering
\includegraphics[width=\columnwidth]{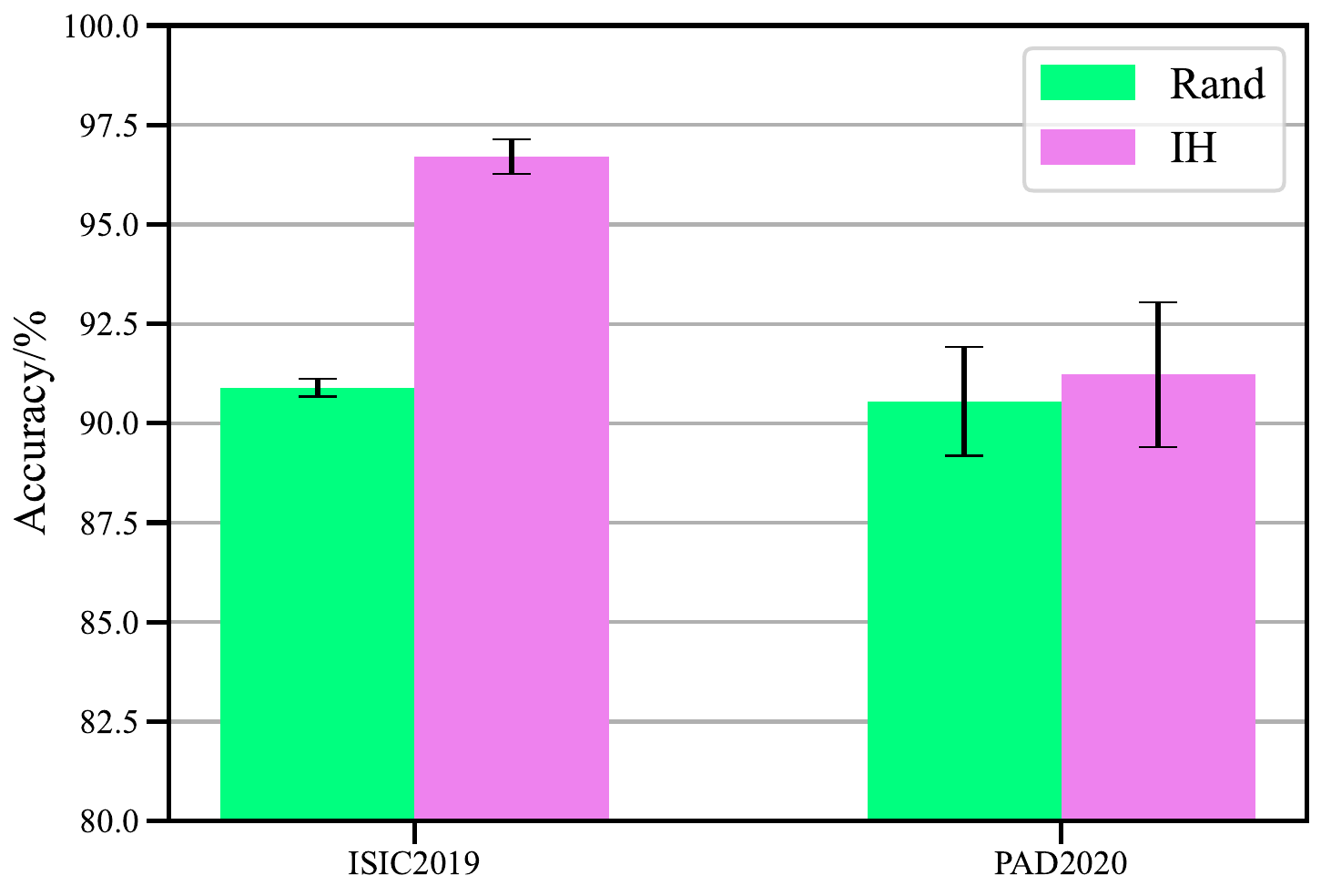}
\caption{\bf Impact of different data balance methods.}
\label{fig:top1datab}
\end{figure}

Fig. \ref{fig:top1datab} elucidates the accuracy of different data balance methods on ISIC2019 and PAD2020. Firstly, the accuracy of IH (instance hardness) is higher than that of Rand (randomised sampling) on both datasets, which means IH can more effectively sample images than Rand. Secondly, IH exceeds 5.9\% and 0.7\% accuracy than Rand in ISIC2019 and PAD2020, respectively. The diminishing improvement in the PAD dataset could be caused by the low imbalance ratio of PAD compared to the ISIC dataset (16.3 for PAD vs. 53.9 for ISIC). The effects of alleviating negative influence on classification are more obvious for the dataset with the higher imbalance ratio. Thirdly, the error bar of ISIC2019 is shorter than PAD2020, indicating that HierAttn is more robust on ISIC2019 than PAD2020.

\subsubsection{Attention blocks in DWSConv}
Table \ref{tab:mechanism} shows the number of parameters, inference time and accuracy of models using different attention mechanisms in DWSConv. It can be discovered that without the attention mechanism [see Fig. \ref{fig:scattn} (a)], the network only achieved only 96.20\% accuracy. However, HierAttn with SEAttn or SCAttn attains at least 0.45\% more accuracy than without using the attention mechanism in DWSConv. Although the accuracy of HierAttn with SCAttn is only 0.05\% more than HierAttn with SEAttn, HierAttn with SCAttn is faster for inference (averaging 1000 iterations) with a smaller number of parameters than HierAttn with SEAttn.

\begin{table}[!h]
\caption{\label{tab:mechanism} {\bf Impact of different attention mechanisms in DWSConv.}}
\centering
\begin{tabular}{|M{1.6cm}|M{2.3cm}|M{1.5cm}|M{1.7cm}|} \hline
{\bf Attention} & \multirow{2}{*}{\bf \# Parameters/M} & {\bf Inference} & \multirow{2}{*}{\bf Accuracy/\%} \\ 
{\bf mechanism} &  & {\bf  time/ms} & \\ \hline
- & { 2.14} & {\bf 1.551}  & 96.20\\ \hline
SEAttn & 2.17 & 1.663 & 96.65\\ \hline
{\bf SCAttn} & {\bf 2.14} & 1.658 & {\bf 96.70}\\ \hline
\end{tabular}
\end{table}

\subsubsection{Transfer learning warm-up}
We consider that the models are unstable after partially transferring tunable parameters and randomly initialising those layers without transferring. Hence, we apply a warm-up technique to alleviate the influence of random weight initialisation for transferred layers. We freeze the transferred layers by stopping gradient backpropagation in the first 30 epochs. In these 30 epochs, only those layers without transferring can update their parameters. Table \ref{tab:accabla} shows that the performance of the HierAttn\_s with the transfer learning warm-up improves by 0.42\% on the IHISIC2019 dataset.

\subsubsection{Stochastic depth} 
Stochastic depth, also regarded as "layer dropout", is implemented in each layer with a skip connection in HierAttn. Typically, it is in all SCADW blocks with a stride of 1 and all CTH blocks. Table \ref{tab:accabla} demonstrates that the stochastic depth effectively enhances the performance of HierAttn\_s by 0.65\%. Note that even without this stochastic depth, the performance of HierAttn\_s delivers similar or better results than SOTA mobile models. 

\subsubsection{Skip connection of CTH block}
We add a skip connection link in the CTH (convolution-transformer hybrid) block with a stride of 1. Moreover, we also apply stochastic depth to those modules with skip connections. Thus, we can reuse the lower lever feature and alleviate the gradient descent. Table \ref{tab:accabla} shows a 0.23\% improvement in the performance of HierAttn\_s with the skip connection. It demonstrates that stochastic depth can reduce the negative influence of adding two distinctive features in the skip connection of the CTH block.

\begin{table}[!h]
\caption{\label{tab:accabla} {\bf Impact on the accuracy under different ablation conditions.}}
\centering
\begin{tabular}{|M{1.4cm}|M{1.7cm}|M{1.7cm}|M{1.7cm}|} \hline
{\bf Condition} & {\bf Warm-up} & {\bf Stochastic depth} & {\bf Skip connection}\\ \hline
Without & 96.28 & 96.05 & 96.47\\ \hline
With & {\bf 96.70} & {\bf 96.70} & {\bf 96.70}\\ \hline
\end{tabular}
\end{table}

\section{Discussion}\label{sec:disc}
From Table \ref{tab:mechanism}, we can also discover that even without attention after depthwise convolution, HierAttn still achieves the most considerable accuracy, 96.20\%, than other SOTA lightweight models on the ISIC2019 validation set. It suggests that branch attention could be preferable to obtaining information on critical stages. It also shows a high potential to use branch attention as a general method for improving the performance of different models. Applying branch attention in a model with more layers, e.g., 300, can prevent the gradient descent in backpropagation.

Additional experiments were also conducted to investigate the inference time, and the results demonstrate that HierAttn\_xs consumes lower than 1 ms for processing one image, 12\% faster than MnasNet1.0’s and just 9\% slower than MobileNetV2’s. Although the self-attention and branch attention mechanisms partially increase the network complexity, the HierAttn\_xs model is still among the middle level in the inference time, with an acceptable sub-millisecond time cost for self-examination or clinical diagnosis.

We also proposed a larger version of HierAttn, HierAttn\_m, with 5.44 M parameters. However, we have not trained or validated HierAttn\_m due to limited computation resources. We believe it also has appreciable potential to detect skin lesions in large hospitals or clinical centres that can support heavy computing in routine examinations.

Mobile application for skin cancer diagnosis allows dermatologists to perform point-of-care testing. Moreover, possible patients can carry out further detection by utilising mobile applications while doing regular self-exam. HierAttn has a statistically close speed to classic mobile networks, which shows great potential to be developed on mobile devices. If the skin lesion is recognised as MEL, BCC, ACK, SCC or VASC with more than 50\% possibility, users are suggested to go to a clinic or hospital to perform further diagnosis. Otherwise, it is more likely that the detected area of the skin is healthy. We expect that HierAttn can be deployed on the mobile phone to assist ordinary people in performing regular self-check in the future. 

\section{Conclusion}\label{sec:con}
In this paper, we propose a HierAttn network consisting of stage attention, branch attention, and SCAttn for skin lesions diagnosis. Stage attention consists of a SCADW block for downsizing feature maps and a CTH block for effectively learning the local and global representations. Branch attention applies hierarchical pooling after each stage attention to learn local and global representations and improve the feature interactions. SCAttn directly extracts global features by using only global average pooling without operating channel-wise information redundantly. With these novel modules, HierAttn can achieve better skin lesion classification results, 96.70\% accuracy and 0.9972 AUC on ISIC2019 and 91.22\% accuracy and 0.98816 AUC on PAD2020 validation set, than other SOTA mobile networks. Moreover, HierAttn is the smallest model among SOTA mobile networks, which shows great potential to be deployed in mobile devices for broader impacts on the general public.

\section*{Declaration of Competing Interest}
The authors declare that they have no known competing financial interests or personal relationships that could have appeared to influence the work reported in this paper.

\section*{CRediT authorship contribution statement}
{\bf Wei Dai}: Conceptualization, Methodology, Validation, Writing. {\bf Rui Liu}: Conceptualization, Methodology. {\bf Tianyi Wu}: Conceptualization, Validation. {\bf Min Wang}: Conceptualization. {\bf Jianqin Yin}: Writing. {\bf Jun Liu}: Conceptualization, Writing, Supervision.

\section*{Acknowledgments}
This work was supported by the Research Grant Council (RGC) of Hong Kong under Grant 11212321 and Grant ECS-21212720, Guangdong Province Basic and Applied Basic Research Fund Project 2019A1515110175, and the Science and Technology Innovation Committee of Shenzhen under Grant Type-C 2022/86.

\section*{Supplementary Material}
Supplementary material associated with this article can be found in the attachment file "Supplement.pdf".

\bibliographystyle{model2-names.bst}\biboptions{authoryear}
\bibliography{refs}

\end{document}